\documentclass[12pt]{article}
\usepackage{subfigure}
\usepackage{epsfig}

\input epsf
\def\esp{\mathrm{e}}
\def\beq{\begin{eqnarray}}
\def\eeq{\end{eqnarray}}
\def\k{{\bf{k}}}
\def\d{{\rm{d}}}

\usepackage{amsmath}
\usepackage{amsfonts}
\usepackage{amssymb}

\begin{document}

\begin{titlepage}

\thispagestyle{empty}

\begin{flushright}
{NYU-TH-09/04/16}
\end{flushright}
\vskip 0.9cm

\centerline{\Large \bf  Vortex Structure in Charged Condensate}

\vskip 0.7cm
\centerline{\large Gregory Gabadadze and Rachel A. Rosen}
\vskip 0.3cm
\centerline{\em Center for Cosmology and Particle Physics,}
\centerline{\em Department of Physics, New York University,}
\centerline{\em New York, NY  10003, USA}

\vskip 1.9cm

\begin{abstract}

We study magnetic fields in the charged condensate that we have previously argued should be present  in helium-core white dwarf stars.  We show that below a certain critical value the magnetic field is entirely expelled from the condensate, while for larger values it penetrates the condensate within flux-tubes that are similar to Abrikosov vortex lines; yet higher fields lead to the disruption of the condensate.  We find the solution for the vortex lines in both relativistic and nonrelativistic theories that exhibit the charged condensation.  We calculate the energy density of the vortex solution and the values of the critical magnetic fields. The minimum magnetic field required for vortices to penetrate the helium white dwarf cores ranges from roughly $10^7$ to $10^9$ Gauss.  Fields of this strength have been observed in white dwarfs. We also calculate the London magnetic field due to the rotation of a dwarf star and show that its value is rather small.

\end{abstract}

\vspace{0.5in}

\end{titlepage}

\newpage

\section {Introduction and Summary}

Consider a system of electrically charged massive scalars and massive fermions of the opposite charge at high densities.  When the interparticle separation is small, or the temperature of the system is high,  neutral fermion-boson bound states  
will be unable to form.  At high temperatures the equilibrium state of the system 
is a plasma.    As the system cools below a certain critical temperature, 
the energy of the Coulomb interactions will significantly exceed the
thermal energy,  and the ionized system can crystallize.  
However, in certain cases the de Broglie wavelengths of the scalars begin to overlap before the 
crystallization temperature is reached.  In this case, instead of crystallizing, 
the quantum-mechanical  probabilistic  ``attraction'' of the bosons can force the scalars 
to undergo condensation into a zero-momentum macroscopic state of large occupation number. 
The scalars minimize their kinetic energy, while phonons can not be thermally excited since the phonon mass gap -- produced 
by this condensate --  happens to be greater than the corresponding temperature.  
Therefore,  after the phase transition all the thermal energy is
stored in the near-the-Fermi-surface gapless excitations of quasi-fermions. 
We refer to this state as a charged condensate \cite{GGRR1,GGRR3,GGDP}. 

This condensation mechanism is different from that of the abelian Higgs model, or equivalently, 
the relativistic Ginzburg-Landau theory of superconductivity --  the scalar field in our case 
has a conventional positive-sign mass term.  
A nonzero expectation value for the electric potential $\langle A_0 \rangle$, or a nonzero 
chemical potential for the scalars,  plays the role of the tachyonic mass, enabling the scalar 
field to acquire a vacuum expectation value \cite {Linde}.  The condensation mechanism 
is generic: 
the electromagnetic interaction can easily be generalized to any $U(1)$ abelian interaction, 
and the scalar field could be a fundamental field or a composite state.

In this work we argue that the charged condensate has properties somewhat  similar to type 
II superconductors. In particular,  we show that it  can admit solutions that are 
similar to the Abrikosov  vortices \cite{Abri}, originally  found in the Ginzburg-Landau 
model of superconductivity,  and later recovered in the relativistic 
abelian Higgs model in \cite{NO}.  The vortex solution is a topologically stable configuration, characterized by a nonzero winding number of the phase of the complex scalar field.
Asymptotically, the scalar field is given by $\phi \sim v \esp^{i \theta}$, where $v$ is 
the vacuum expectation value of the field and $\theta$ is the azimuthal coordinate.  
Like the Abrikosov vortex, the charged condensate vortex carries a quantized magnetic 
flux.  The vortex solution has a higher energy density than the pure condensate solution.  
However, in the presence of a sufficiently high external magnetic field, it becomes 
energetically favorable for the charged condensate to form vortices.

The obtained vortex line solution exhibits the following structure: it has a  
narrow cylindrical core where the scalar field changes 
significantly from a nonzero to a zero value; this core is surrounded by a 
broad halo in which the  magnetic flux is confined. The width of the latter region 
is determined by  the penetration depth (i.e., the photon Compton wavelength). 
We refer to the system of the core and the halo as  the flux-tube, or the vortex line. 
This structure is similar to that of the Abrikosov solution.  However, unlike the latter, 
our solution also carries a  profile of the electrostatic potential within the 
halo, while this potential is exponentially small outside of the halo (i.e., the 
flux tube is charge neutral).  Hence, in terms of differential equations, 
one has to solve three coupled equations  instead of the 
two required in the Abrikosov \cite{Abri},  or Nielsen-Olesen cases  \cite{NO}.

One  application of the charged condensate is to helium-core  white dwarf stars.  
The cores of these dwarfs are composed of a highly dense system of helium-4 nuclei and 
electrons.  At high temperatures, the electrons and nuclei form a plasma.  While white dwarf 
stars composed of carbon, oxygen or heavier elements are expected to crystallize as they cool
\cite {Ruderman}, it was argued in Refs. \cite{GGRR3,GGDP} that the helium-4 nuclei would 
instead form  the charged condensate, as the condensation temperature in the helium dwarfs 
is higher than the temperature at which the system would crystallize. This 
transition dramatically affects the cooling history of the helium-core white dwarfs. 
In fact, they cool faster; as a result, the luminosity function exhibits a sharp 
drop-off below the condensation temperature \cite{GGDP}. 
Such a termination in the luminosity function may have  already been observed 
in a sequence of the 24 helium-core white dwarf candidates seen in NGC 6397 \cite {6397}. 

The above conclusions were obtained by considering white dwarfs that are not magnetized.  Magnetized helium-core dwarf stars are also believed to exist; in magnetized white dwarf stars, surface magnetic fields have been detected ranging from $10^3$ to $10^9$ Gauss.  If this is the case, it is important to know how the presence of a magnetic field would affect the above-described properties of the charged condensate.

In analogy with type II superconductors, we would expect an external magnetic field to be 
entirely expelled from the charged condensate below a certain critical value of the field $H_{c1}$.  Above this value, however, we'd expect to have a mixed phase in which the magnetic field penetrates the charged condensate only in the form of the Abrikosov-like vortices. Finally, above a certain $H_{c2}> H_{c1}$ the magnetic field should entirely destroy the charged condensate.  

Indeed, this is the pattern that we find in the present work. The corresponding values  of the critical magnetic fields in the interior of  helium-core white dwarfs are: $H_{c1}\simeq (10^7- 10^9)$ Gauss, and  $H_{c2} \simeq  (10^{13} - 10^{15})$  Gauss, while in  between these two scales we find quantized magnetic vortices permeating the bulk of helium-core dwarf stars.  Hence in most of the magnetized helium white dwarfs the magnetic field will be expelled from  the core in which the charged condensation has taken place.  It's only the highly magnetized dwarfs, with  fields $\sim (10^7-  10^9)$ Gauss, that can admit vortices.  Also, since  the value of $H_{c2} \simeq (10^{13}- 10^{15})$ Gauss is much higher that any  magnetic field that may be present in dwarf stars, there will be no disruption of the charged condensate due to magnetic fields. 

The presence of the magnetic field would decrease  the critical temperature at which the 
charged condensation would take place. However, this decrease 
will be significant only for the fields $H$ close to the critical 
ones. For smaller fields,  the change in the value of the critical temperature 
will be small $\Delta T_c\sim - T_c (H/H_c)$.

The organization of this paper is as follows:  In section 2 we briefly describe the 
condensation mechanism for a generic system of scalars and fermions in the context of 
relativistic field theory.  In section 3 we fix the phase of the scalar field to be of 
the vortex-type and solve the corresponding equations of motion.  
We compare our solutions to those found in the abelian Higgs model.  
In section 4 we consider corrections to our solutions due to the dynamics 
of the fermions.  In section 5 we consider the effects of an external magnetic 
field on the charged condensate and determine the magnitude of the external 
field for which it becomes energetically favorable to form vortices.  
In section 6 we treat specifically the case of helium-4 nuclei and electrons.  
We describe the system in the context of a non-relativistic low energy effective 
field theory, rather than the relativistic field theory used in the previous 
sections.  We discuss the applicability of the vortex solutions found in section 
4 to the helium-4 nuclei and electron system.  We also consider the effect of a 
constant rotation on the condensate of helium-4 nuclei.

\section {Charged condensate: relativistic Lagrangian}
We start by considering a generic, highly dense system of charged, massive scalars and oppositely charged fermions.  We assign a charge of $+2e$ to the scalars and $-e$ to the fermions in anticipation of the helium-4 nuclei and electron system to be discussed later in section 6.  However, for now we keep our considerations general.  The scalar field considered below could be any 
fundamental scalar field, possibly originating in beyond-the-standard-model physics.  
Our conclusions are independent of the specific charge assignment.

The scalar condensate is described by the order parameter $\phi$.  A nonzero vacuum expectation value of $\phi$ implies that the scalars are in the condensate phase.  Here we adopt a relativistic Lorentz-invariant Lagrangian which contains the charged scalar field $\phi$, and the photon field $A_\mu$.  The fermion current is given by $J_\mu$.
\beq
\label{lagr0}
{\cal{L}} = -\tfrac{1}{4} F_{\mu\nu}^2 +  \vert D_{\mu} \phi \vert^2
 - m_H^2 \phi^{\ast} \phi -e A^{\mu} J_{\mu} \, .
\eeq
The covariant derivative for the scalars is defined as $D_\mu \equiv \partial_\mu-2 i e A_\mu $.

The above Lagrangian could also contain a quartic interaction $\lambda (\phi^* \phi)^2$.  In the case that $m_H^3 \gg \lambda J_0$, this term will not alter significantly our results.  Other possible terms including a Yukawa interaction between fermions and scalars were considered in Ref. \cite{GGRR1}. For now we treat the fermions as a fixed, constant background charge density $J_\mu = J_0 \delta_{\mu 0}$.  We will relax this assumption in section 4 and consider effects due to the dynamics of the fermions and quantum loops of relativistic fermions.  For simplicity we take the system to be at zero temperature (for some discussions of finite temperature effects see, \cite {GGDP}).

Because the system has a conserved scalar current, we can associate with it a chemical potential $\mu_s$.  For the Hamiltonian density, the inclusion of a chemical potential for the scalars results in the shift ${\cal{H}} \rightarrow {\cal{H}}' = {\cal{H}}-\mu_s J_0 ^{\rm{scalar}} $, where $J_0 ^{\rm{scalar}} \equiv -i [(D_0 \Phi)^* \Phi-\Phi^* (D_0 \Phi)]$ is the time component of the conserved scalar current.  For the Lagrangian density this shift can be written as a shift in the covariant derivative $D_\mu \rightarrow D_\mu' = D_\mu-i\mu_s \delta_{\mu0}$.  In what follows primed variables ${\cal{H}}'$, ${\cal{L}}'$ will refer to those variables which include a nonzero chemical potential for the scalars.

The complex order parameter $\phi$ can be written in terms of a modulus and a phase $\phi = \tfrac{1}{\sqrt{2}} \sigma\, \esp^{i \alpha}$.  In these variables the Lagrangian density becomes 
\beq
\label{lagr}
{\cal{L}}'=-\tfrac{1}{4} F_{\mu\nu}^2 + 
\tfrac{1}{2}(\partial_{\mu}\sigma)^2+
\tfrac{1}{2}(2eA_\mu+\mu_s \delta_{\mu 0}-\partial_\mu \alpha)^2 \, \sigma^2- 
\tfrac{1}{2} m_H^2 \, \sigma^2 - e A^{\mu} J_{\mu} \, .
\eeq
From this form of the Lagrangian it is evident that a nonzero expectation value for $A_0$ 
or a nonzero chemical potential $\mu_s$ act as a tachyonic mass for the scalars \cite {Linde}.  
In particular, when $\langle 2e A_0 \rangle+\mu_s = m_H$, the scalar field condenses, 
as we shall now show.

Varying the Lagrangian with respect to $A_\mu$ and $\sigma$ gives the following equations of motion:
\beq
\label{eom1}
-\partial^\mu F_{\mu \nu}& =& 2e(2eA_\nu+\mu_s \delta_{\nu 0}-\partial_\nu \alpha) \sigma^2-eJ_\nu \, ,  \\
\label{eom2}
\Box \, \sigma &= &[(2eA_\mu+\mu_s \delta_{\mu 0}-\partial_\mu \alpha)^2-m_H^2] \, \sigma \, .
\eeq
Varying with respect to $\alpha$ gives the conservation of the scalar current:
\beq
\label{J0scalar}
\partial^\mu J_\mu^{\rm{scalar}} = \partial^\mu \left[(2eA_\mu+\mu_s \delta_{\mu 0}-\partial_\mu \alpha) \, \sigma^2 \right] = 0.
\eeq
On equation of motion (\ref{eom1}) this expression is automatically satisfied.

We now work in the unitary gauge where the phase of the scalar field is set to zero: $\alpha =0$.  
Note that this gauge choice is acceptable for a classical description of the condensate, however, 
in subsequent sections we will not be able to choose this gauge for the vortex solution for 
a well-known reason: 
in order to take $\alpha = \theta$ to $0$, the corresponding gauge transformation $A_\mu \rightarrow A_\mu +\partial_\mu \theta$ would be singular at the origin where the vortex core is located. 

In the unitary gauge the equations of motion (\ref{eom1}), (\ref{eom2}) 
have the following static solution with a nonzero expectation value for $\sigma$:
\beq
\label{staticsol}
\langle 2 e A_0 \rangle+\mu_s= m_H,  ~~~~\langle \sigma \rangle = \sqrt{\frac{J_0}{2m_H}} \, .
\eeq

In the condensate the gauge symmetry is spontaneously broken.  The photon acquires a mass 
\beq
\label{mgamma}
m_\gamma = 2e \, \sqrt{\frac{J_0}{2m_H}} \, .
\eeq
This mechanism of symmetry breaking differs from the abelian Higgs model in that here the scalar field has a conventional positive-sign mass term.  A nonzero expectation value for $A_0$ or a nonzero chemical potential $\mu_s$ act as a tachyonic mass term.  If we consider a system with no net charge then $ \langle A_0 \rangle= 0$.  Then, from expression (\ref{staticsol}), in order for the scalar field to condense the chemical potential must satisfy $\mu_s =m_H$.  The bulk of the condensate is always electrically neutral, the scalar charge density exactly canceling the fermion charge density:  $2eJ_0^{\rm scalar} = (\langle 2 e A_0 \rangle+\mu_s) \langle \sigma^2 \rangle = e J_0$.

\section{Vortices in charged condensate}

To find the charged condensate solution (\ref{staticsol}), we fixed the phase of the scalar field to zero.  We now consider a configuration where the phase is not set to zero, nor can it be set to zero everywhere by a non-singular gauge transformation.  The requirement that the scalar field be single-valued everywhere is satisfied by demanding that the change in phase around a closed loop be an integer multiple of $2 \pi$.  In a system with cylindrical symmetry, this is satisfied by setting $\alpha = n \theta$, where $\theta$ is the azimuthal coordinate and $n$ is an integer.  
This phase can be removed everywhere by a gauge transformation $A_\mu \rightarrow A_\mu +\partial_\mu (n \theta)$, except at the origin where the scalar VEV goes to zero   and the gauge transformation would be singular. The solutions of the 
equations of motion (\ref{eom1}), (\ref{eom2}) where the phase is 
fixed to $\alpha = n \theta$ 
are vortex-type solutions.

Hence, $\sigma = 0$ at  the origin $r=0$ (where $r$ is the 2D radial coordinate).  
Far from the origin however, we expect the solutions to recover the condensate 
solutions (\ref{staticsol}).  At large $r$ then, the gauge field takes the form 
$2eA_j \rightarrow \partial_j \alpha$, or equivalently $A_\theta \rightarrow n/(2er)$.  
From this form of the vector potential, it follows that this configuration has a 
quantized magnetic flux ${\it \Phi}$ that is related to the integral of $A_j$ 
around a closed loop at infinity:
\beq
\label{flux}
{\it \Phi} = \oint {\bf{A \cdot}} \d {\bf{l}}= \oint A_\theta r \d \theta  = \frac{2 \pi n}{2e}\, .
\eeq
The magnetic flux is quantized in units of $n$.  The quantization of flux implies the stability of the vortex configuration, although it may be possible for a high $n$ vortex to decay into multiple vortices of smaller $n$.

To solve the equations of motion (\ref{eom1}), (\ref{eom2}) for the vortex configuration we switch notation to dimensionless variables.  The resulting equations are governed by a single parameter $\kappa$, the ratio of the mass of the scalar to the mass of the photon in the condensate: $\kappa = m_H/m_\gamma$.  This parameter $\kappa$ is the equivalent to the Ginzburg-Landau parameter in the theory of superconductors which gives the ratio of the penetration depth to the coherence length.  For the helium white dwarf star, if we take the mass of the helium-4 nuclei to be roughly $m_H =3.7$ GeV and the electron density to be $J_0 \sim (0.15-0.5 {\rm{~MeV}})^3$, then we have $\kappa \sim 10^6$.  In our derivations below we frequently take the large $\kappa$ limit.

We define $x \equiv m_\gamma r$, set $A_r = A_z = 0$ and $\mu_s = m_H$, and perform the following change of variables:
\beq
\label{changevar}
m_\gamma A(x) & \equiv& 2exA_\theta(x) \, , \\
m_\gamma F(x) & \equiv& 2e \sigma(x) \, , \\
m_H B(x)& \equiv& \mu_s+2eA_0(x)\, .
\eeq
In terms of these new variables equations of motion (\ref{eom1}), (\ref{eom2}) become
\beq
\label{eomA}
x \, \frac{d}{dx}\left(\frac{1}{x}\frac{dA}{dx}\right)&=&F^2 (A-1) \, , \\
\label{eomF}
-\frac{1}{x}\frac{d}{dx}\left(x\frac{dF}{dx}\right)&=&\left[\kappa^2(B^2-1)-\frac{n^2}{x^2}(A-1)^2\right]F \, , \\
\label{eomB}
\frac{1}{x}\frac{d}{dx}\left(x\frac{dB}{dx}\right)&=&F^2 B -1 \, .
\eeq
The boundary conditions are set by requiring that the solutions asymptote to the condensate solutions for large $r$, while for $r=0$ we have $A_\theta = \sigma = dA_0/dr = 0$:
\beq
\label{bc}
\begin{array}{llll}
{\rm{For }} ~x \rightarrow 0: & A(x) \rightarrow 0, &  F(x) \rightarrow 0, & \frac{dB}{dx}  \rightarrow 0 \, . \\
{\rm{For }} ~x \rightarrow \infty: & A(x) \rightarrow 1, & F(x) \rightarrow 1, & B(x) \rightarrow 1\, . \\
\end{array}
\eeq

\vspace{0.5cm}

We can compare these expressions to those obtained in the usual abelian Higgs model.  Suppose that instead of Lagrangian (\ref{lagr}) we had the abelian Higgs Lagrangian:
\beq
\label{LAH}
{\cal L}^{\rm AH} =-\frac{1}{4} F_{\mu\nu}^2 + 
\frac{1}{2}(\partial_{\mu}\sigma)^2+
\frac{1}{2}(2eA_\mu-\partial_\mu \alpha)^2 \, \sigma^2- 
\frac{\lambda}{4} (\sigma^2 - v^2)^2 \, .
\eeq
Using the same change of variables as above and defining $m_H^{\rm AH} = \sqrt{\lambda} v$, $m_\gamma^{\rm AH} = 2ev$, the equations of motion are:
\beq
\label{eomAAH}
x \, \frac{d}{dx}\left(\frac{1}{x}\frac{dA}{dx}\right)&=&F^2 (A-1) \, , \\
\label{eomFAH}
-\frac{1}{x}\frac{d}{dx}\left(x\frac{dF}{dx}\right)&=&\left[\kappa^2(1-F^2)-\frac{n^2}{x^2}(A-1)^2\right]F \, .
\eeq
The equation of motion for the vector potential, expressed via $A$, is the same as in the charged condensate model.  In the equation for the scalar field, the $\sigma^4$ term in the abelian Higgs model gets replaced in the charged condensate model by a term that depends on the electric potential.  In addition, in the charged condensate equations the electric potential is generally not zero and not constant and has its own equation to satisfy.

\vspace{0.5cm}

Let us first examine the asymptotic behavior of the solutions to the condensate equations for $x \rightarrow \infty$.  Far from the origin we expect the fields to be very close to their condensate values.  Then, on the r.h.s. of equation (\ref{eomA}), it follows that $A(x)-1 \equiv a(x)$ is very small.  If we consider this equation only to first order in small fields then we can approximate the scalar field on the r.h.s. of (\ref{eomA}) as $F \simeq 1$.  The solution for $A$ that obeys the appropriate boundary conditions is
\beq 
\label{Asol}
A(x)=1-c_a x K_1(x) \, .
\eeq  
Here $c_a$ is a constant to be determined by the matching of the solutions and $K_1(x)$ is the modified Bessel function of the second kind.  In the large $x$ limit this solution for $A$ becomes
\beq
\label{Alarge}
A(x) \rightarrow 1-c_a \sqrt{\frac{\pi x}{2}} {\rm{e}}^{-x} \,.
\eeq

To find the asymptotic behavior of $B$ and $F$ we expand these fields in terms of perturbations above the condensate values, $B(x)=1+b(x)$ and $F(x)=1+f(x)$, and we assume that $b(x), f(x) \ll1$.  We then substitute these expressions as well as expression (\ref{Alarge}) into the equations for $B$ and $F$ and keep only terms linear in the perturbations $b(x)$ and $f(x)$.  These two equations can be combined to obtain a fourth order differential equation for $b(x)$.  Using the ansatz $b(x) = c_b \, x^s \, {\rm{e}}^{-kx}$ where $c_b$, $s$, and $k$ are as yet undetermined constants, we can find the particular and homogeneous solutions for $b(x)$ in the large $x$ limit.  We also take $c_a \simeq 1$ which we will justify later.  For the particular solution we find
\beq
\label{bp}
b_p(x) = \frac{\pi n^2}{4 (\kappa^2+3)}\frac{{\rm{e}}^{-2x}}{x} \, .
\eeq
For the homogeneous solution we deduce $s=-1/2$ and $k^2 = (1\pm \sqrt{1-16 \kappa^2})/2$.  In the limit that $\kappa$ is very large, the solution becomes
\beq
\label{bh}
b_h(x) = \frac{{\rm{e}}^{-\sqrt{\kappa}x}}{\sqrt{x}} \left[c_1 \sin(\sqrt{\kappa}x) +c_2 \cos(\sqrt{\kappa}x)\right] \, ,
\eeq
with  some constants $c_1$ and $c_2$.  The complete solution is then
\beq
\label{Btot}
B(x) = 1 + \frac{\pi n^2}{4(\kappa^2+3)}\frac{{\rm{e}}^{-2x}}{x} + \frac{{\rm{e}}^{-\sqrt{\kappa}x}}{\sqrt{x}} \left[c_1 \sin(\sqrt{\kappa}x) +c_2 \cos(\sqrt{\kappa}x)\right]\, .
\eeq
The solution for $F(x)$ can be found once $B(x)$ is known:
\beq
\label{Ftot}
F(x) = 1 + \frac{3\pi n^2}{8(\kappa^2+3)}\frac{{\rm{e}}^{-2x}}{x} - \frac{\kappa {\rm{e}}^{-\sqrt{\kappa}x}}{\sqrt{x}} \left[c_1 \cos(\sqrt{\kappa}x) -c_2 \sin(\sqrt{\kappa}x)\right]\, .
\eeq
Here $c_1$ and $c_2$ are the same integration constants that appear in the expression for $B(x)$.  

As $\kappa$ is large, the second term in the above expressions for $B$ and $F$ dominates 
the asymptotic behavior.  For $x\rightarrow \infty $ we have
\beq
\label{Basym}
B(x) \rightarrow 1 + \frac{\pi n^2}{4(\kappa^2+3)}\frac{{\rm{e}}^{-2x}}{x} \, , \\
\label{Fasym}
F(x) \rightarrow 1 + \frac{3\pi n^2}{8(\kappa^2+3)}\frac{{\rm{e}}^{-2x}}{x} \, .
\eeq
The asymptotic behavior for the vector potential and the scalar field are similar to that for the abelian Higgs model:
\beq
\label{AFasymHiggs}
A^{\rm AH}(x)  = 1-c_a \sqrt{\frac{\pi x}{2}} {\rm{e}}^{-x} \, , ~~~~~~
F^{\rm AH}(x) = 1 - \frac{c_a^2 \pi n^2}{4(\kappa^2-2)}\frac{{\rm{e}}^{-2x}}{x} +c_f \frac{\esp^{-\sqrt{2}\kappa x}}{\sqrt{x}}\, . 
\eeq
The vector potential $A$, and thus the magnetic field, are the same in both the charged condensate and abelian Higgs models.  The asymptotic behavior of the scalar field in the abelian Higgs model in the large $\kappa$ limit is dominated by the ${\rm{e}}^{-2x}$ term, as in the charged condensate model.

Notably, this is not the asymptotic behavior for the Abrikosov vortex given in 
the Nielsen-Olsen paper \cite{NO}.  This discrepancy was first pointed out by  
L. Perivolaropoulos in \cite{Peri}.  The incorrect asymptotic behavior is obtained if one similarly expands $A(x)$ as $A(x) = 1+a(x)$ and only keeps terms linear in $a(x)$.  This is because the last term in (\ref{eomF}) and the last term in (\ref{eomFAH}) are quadratic in $a(x)$ and yet, due to the different exponential dependence of the perturbations $a(x)$, $b(x)$ and $f(x)$, these terms can be dominant over terms which are linear in $b(x)$ and $f(x)$.  Linearizing $A$ gives the correct asymptotic behavior of the fields only in the limit that $\kappa$ is small.

The second term in the full expressions for $B$ and $F$ and thus the asymptotic behavior of both fields is due strictly to the presence of a nonzero magnetic field.  In the absence of any magnetic field, the screening of any small perturbation of the fields above their condensate values vanishes as $\esp^{-\sqrt{\kappa}x}=\esp^{-\sqrt{m_H m_\gamma}r}$ (see Ref. \cite{GGRR3}).  

\vspace{0.5cm}

The above asymptotic expressions are valid as long as $x \gg 1$, or equivalently $r \gg 1/m_\gamma$.  We consider now the intermediate region $1/\sqrt{\kappa} \ll x \ll 1$, or equivalently $ 1/M \ll r \ll 1/m_\gamma$ where for convenience we have defined $M \equiv \sqrt{m_H m_\gamma}$.  
At distances much larger than $1/M$ we assume that the scalar field $F$ is still close to its condensate value.  Thus expression (\ref{Asol}) is still valid for $A$.  In this regime then $n^2/x^2 (A-1)^2 \simeq n^2/x^2$.  The equations for $B$ and $F$ become
\beq
\label{eomBF2}
\frac{1}{x}\frac{d}{dx}\left[x\frac{dB}{dx}\right]=F^2 B-1\, , ~~~~~
 \frac{1}{\kappa^2}\left[\frac{1}{x}\frac{d}{dx}\left(x\frac{dF}{dx}\right)-\frac{n^2}{x^2}F\right]=(1-B^2)F\, .
\eeq
The solutions are straightforward to find:
\beq
\label{BFsol}
B(x)=\left(1+\frac{n^2}{\kappa^2 x^2}\right)^{1/2}\,,  ~~~~
F(x)=\left(1-\frac{n^2}{2\kappa^2 x^2}+\frac{2n^2}{\kappa^2 x^4}\right)^{1/2} \, .
\eeq
Alternatively, 
we can once again expand $B$ and $F$ above their condensate values, $B(x)=1+b(x)$ and $F(x)=1+f(x)$, and solve for $b(x)$ and $f(x)$.   The homogeneous solutions for $b(x)$ and $f(x)$ are the same as those given above with the same coefficients $c_1$ and $c_2$.  Solving for the particular solutions gives the full solutions in the linearized approximation:
\beq
\label{Blin}
B(x)& =& 1 + \frac{n^2}{2\kappa^2 x^2}+ \frac{{\rm{e}}^{-\sqrt{\kappa}x}}{\sqrt{x}} \left[c_1 \sin(\sqrt{\kappa}x) +c_2 \cos(\sqrt{\kappa}x)\right]\, , \\
\label{Flin}
F(x)& =& 1 - \frac{n^2}{4 \kappa^2x^2}+ \frac{n^2}{\kappa^2x^4}- \frac{\kappa {\rm{e}}^{-\sqrt{\kappa}x}}{\sqrt{x}} \left[c_1 \cos(\sqrt{\kappa}x) -c_2 \sin(\sqrt{\kappa}x)\right]\, .
\eeq
The coefficients $c_1$ and $c_2$  are needed to perform the matching.  However, as we'll see below,  these coefficients will turn out not to be exponentially large, and hence
solutions (\ref {Blin}) and (\ref {Flin}) approximate well the solutions  in (\ref {BFsol}).

The approximations made to find both the homogeneous and particular solutions break 
down as $x$ approaches $1/\sqrt{\kappa}$.  Moreover, $f(x)$ becomes of order 
1 at $x \sim 1/\sqrt{\kappa}$ and thus the linear approximation in general 
no longer holds below this scale.

\vspace{0.5cm}

Finally, we'd like to  solve in the $r \rightarrow 0$ limit.   Before we do so, however, 
we emphasize that validity of this procedure needs some justification.  
The interparticle separation is given by $d \propto J_0^{-1/3}$.  
This corresponds to $x \propto 1/\kappa^{1/3}$.  
At distances shorter than this $x$ we expect that an effective field theory 
would break down and thus it would make little physical sense to solve 
the equations (\ref{eomA}), (\ref{eomF}) and (\ref{eomB}) in this regime.  
Moreover, the scale $1/M$ is typically shorter than  the interparticle separation 
$d$, hence, particles at these scales cannot in general be modeled by  
a smooth distribution. 

However, both fermions and bosons are in a condensate state in which the location of individual 
particles has uncertainties much greater than  the interparticle separation. Hence the 
latter notion loses its meaning as a microscopic characteristic of the system.
For this, we'll  still approximate particle  distributions by smooth  functions
all the way down to the scale $\sim 1/M$, which  is a dynamically determined short-distance 
scale at which weakly coupled expansion breaks down \cite {GGRR1}.  As to solving at scales less that $1/M$, we regard this as a purely 
mathematical exercise aimed at finding the matching of the asymptotic solutions for 
the corresponding differential equations for all values of the coordinate $x$. 

Taking $A$, $B$, and $F$ to be series expansions in small $x$ obeying the 
appropriate boundary conditions, the solutions to (\ref{eomA}), (\ref{eomF}) 
and (\ref{eomB}) are 
\beq
\label{A0}
A(x) &=& a_0 x^2-\frac{f_0^2}{8} x^4 \, , \\
\label{B0}
B(x) &=& b_0 -\frac{x^2}{4} \, , \\
\label{F0}
F(x) &=& f_0 \left[x-\frac{1}{8}\left(\kappa^2(b_0^2-1)+2 a_0\right)x^3\right] \, .
\eeq 
For simplicity we have solved for the case that the winding number $n=1$.\footnote{For $n \neq 1$, the leading term in the expansion for $F$ will be $\propto x^n$.  The expression for $B(x)$ remains unchanged and the leading term in the expansion for $A(x) = a_0 x^2$ is also the unchanged.}  The coefficients $a_0$, $f_0$, and $b_0$, as well as the coefficients $c_a$, $c_1$, and $c_2$ can be determined by matching the above solutions to those in the intermediate region, given by (\ref{Asol}), (\ref{Blin}) and (\ref{Flin}).

To determine the physically appropriate matching radius, we first use Gauss's law to 
find the charge of the vortex solution.  The number density of fermions in the 
vortex $J_0$ is constant and is the same as the number density of fermions 
in the normal condensate phase.  We have fixed it so by hand, but will justify this later.  
The scalar number density is given by $\tfrac{1}{2} J_0 BF^2$ and varies as a function of $x$.  
Therefore it is not in general equal to its condensate value $\tfrac{1}{2} J_0$.  
The variation of the scalar number density away from its condensate value can lead to a 
net charge density of the vortex core within the vortex halo.  
In particular, there are two competing effects.  In the intermediate 
region $1/\sqrt{\kappa} \ll x \ll 1$, both $B$ and $F$ are above their condensate 
values, thus the scalar number density is greater than the scalar number density in 
the condensate.  As $x \rightarrow 0$, however, $F \rightarrow 0$ and the scalar number 
density drops to zero, significantly below the condensate value.  The matching radius 
should be chosen so that these two effects combine to give the appropriate 
charge density as determined by Gauss's law.  

From Gauss's law we can calculate the the average charge density of the vortex inside 
radius $x=1$.  As is usually the case, we can determine the net charge enclosed in a region 
knowing only the form of the potential at the boundary of that region.  Equation (\ref{BFsol}) 
gives the potential in the intermediate region independent of matching coefficients $c_1$ and 
$c_2$.  This form of the potential, together with Gauss's law, allows us to calculate the net 
charge  of the vortex at $x=1$ independent of the matching conditions and the $x \rightarrow 0$ 
solutions.

Gauss's law is given by equation (\ref{eom1}):
\beq
\label{Gauss}
\nabla^2 A_0 = 2e (2eA_0+\mu_s-\dot{\alpha})\sigma^2- e J_0 \, .
\eeq
The r.h.s. of the above equation is the charge density.  Integrating both sides of the above expression over the volume of the vortex and dividing by the total volume gives the average charge density inside distance $x$:
\beq
\label{Qenc}
\frac{Q_{\rm{enc}}}{V} = 2eJ_0\frac{1}{x}\frac{dB}{dx} \, .
\eeq
Here $V$ is the volume equal to the length of the vortex times the cross-sectional area and the r.h.s. is evaluated at the boundary of the vortex.  Using expression (\ref{BFsol}) for $B(x)$ at $x=1$, the average charge density inside $x=1$ is $Q_{\rm{enc}}/V=- 2e J_0/\kappa^2$.  The negative sign indicates a dearth of scalars in this region, but, as $\kappa$ is very large, this is a small correction to the overall average charge density of scalars $\propto e J_0$.  To check this result one can likewise use the asymptotic solution for $B$, expression (\ref{Btot}), at $x=1$.  Assuming that the coefficients $c_1$ and $c_2$ are not exponentially large and thus these terms are not dominant in the solution for $B$ at $x=1$, one finds $Q_{\rm{enc}}/V \propto- 2e J_0/\kappa^2$.  This is consistent with the previous result.  Farther out, $B(x)-1$ is exponentially suppressed thus the net charge of the vortex approaches zero as $x$ becomes large.

We can now use this result to determine the matching radius $R$.  Given the smallness of the average charge density found inside $x = 1$, the excess of scalars in the intermediate region of the vortex must cancel the shortage of scalars in the $x \rightarrow 0$ region to great accuracy.  Using expressions (\ref{BFsol}) for $B$ and $F$, it can be shown that this happens when $R \simeq 1/\sqrt{\kappa}$.  Thus we use this as our matching radius $R$ in what follows.

We start by matching the solution for $A(x)$ in (\ref{A0}) and its first derivative with its solution in the intermediate region (\ref{Asol}).  Taking the matching radius to be small, $R \ll 1$, gives:
\beq
\label{aco}
a_0 = - \frac{1}{2} \left[\gamma +\ln\left(\frac{R}{2}\right) \right] \, , ~~~~ c_a =1+\frac{R^2}{4} \, ,
\eeq
where $\gamma$ is the Euler-Mascheroni constant.  As long as $R$ is less than one, $a_0$ is positive.  Moreover, we see that we were justified in taking $c_a \simeq 1$ in our previous calculations.  The magnetic field is given by 
\beq
\label{mag}
H = \frac{m_\gamma^2}{2e} \, \frac{1}{x} \frac{dA}{dx} \, .
\eeq
Near the origin the magnetic field is of order $m_\gamma^2/(2e)$.   For $x > 1/\sqrt{\kappa}$ it is given by $m_\gamma^2 K_0(x)/(2e)$, where $K_0(x)$ is the modified Bessel function.  For $x  \gg 1$, i.e. for $r  \gg 1/m_\gamma$, the magnetic field is exponentially small.

To find the remaining coefficients, we use $a_0$ obtained above and match (\ref{B0}) and (\ref{F0}) and their first derivatives to the appropriate solutions in the intermediate region (\ref{Blin}), (\ref{Flin}).  We now take the matching radius to be $R=1/\sqrt{\kappa}$.  The solution with the lowest energy is one in which the scalar field $F(x)$ is identically zero in the region $x < R$.  The corresponding coefficients are
\beq
\label{bfco}
\begin{array}{cclccl}
\vspace{.2cm} 
b_0 &=&1+\frac{7}{4 \kappa}  \, ,&~~~~c_1& =&\kappa^{-5/4}\, \esp \left(2 \cos(1)+\sin(1)\right) \, , \\ 
f_0 &=&0\, , &~~~~c_2&=&\kappa^{-5/4}\, \esp \left(\cos(1)-2\sin(1)\right) \, .
\end{array}
\eeq
In Fig.\ref{fig} below, the fields are plotted for small $r$ and for $\kappa \sim 10^6$.  The radius $r=1/M$ corresponds to the matching radius $x=R=1/\sqrt{\kappa}$.  The radius $r = d$ denotes the interparticle separation $d = J_0^{-1/3}$.  Unlike the magnetic field, the potential and scalar field approach their condensate values for $x > 1/\sqrt{\kappa}$.  This is in contrast to the abelian Higgs model in which the scalar field is close to its condensate value for $x > 1/\kappa$, i.e. for $r > 1/m_H$.

\begin{figure}[h!]
\begin{center}
\subfigure[Scalar field as a function of radius]{\epsfig{figure=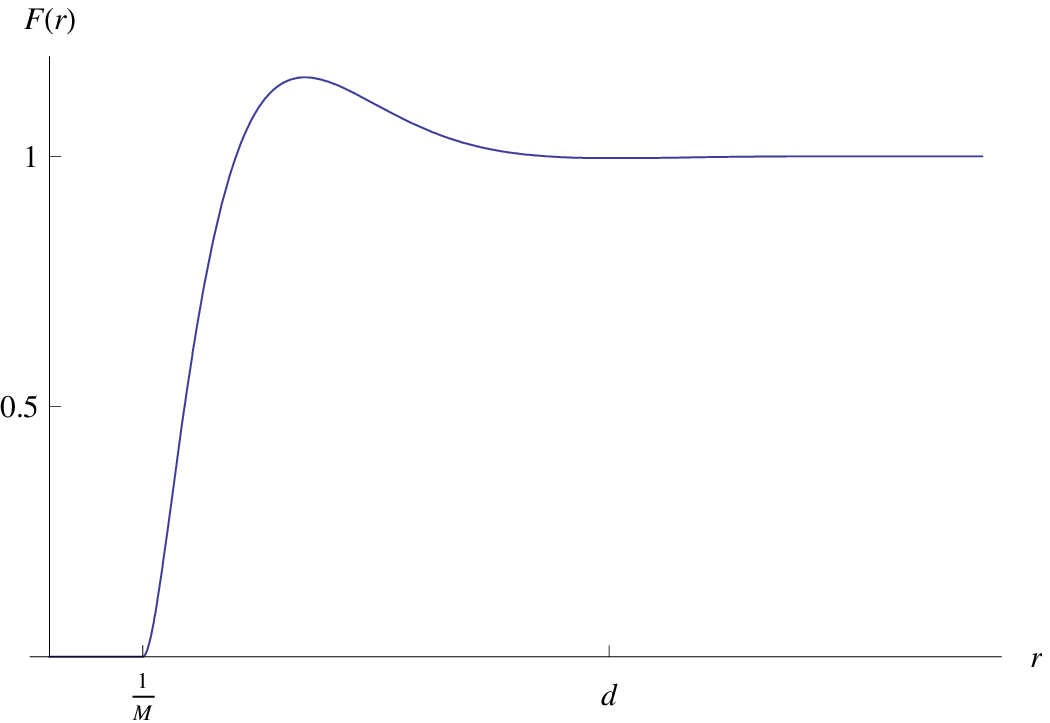,width=.45\textwidth}}
\hskip .15in
\subfigure[Potential as a function or radius]{\epsfig{figure=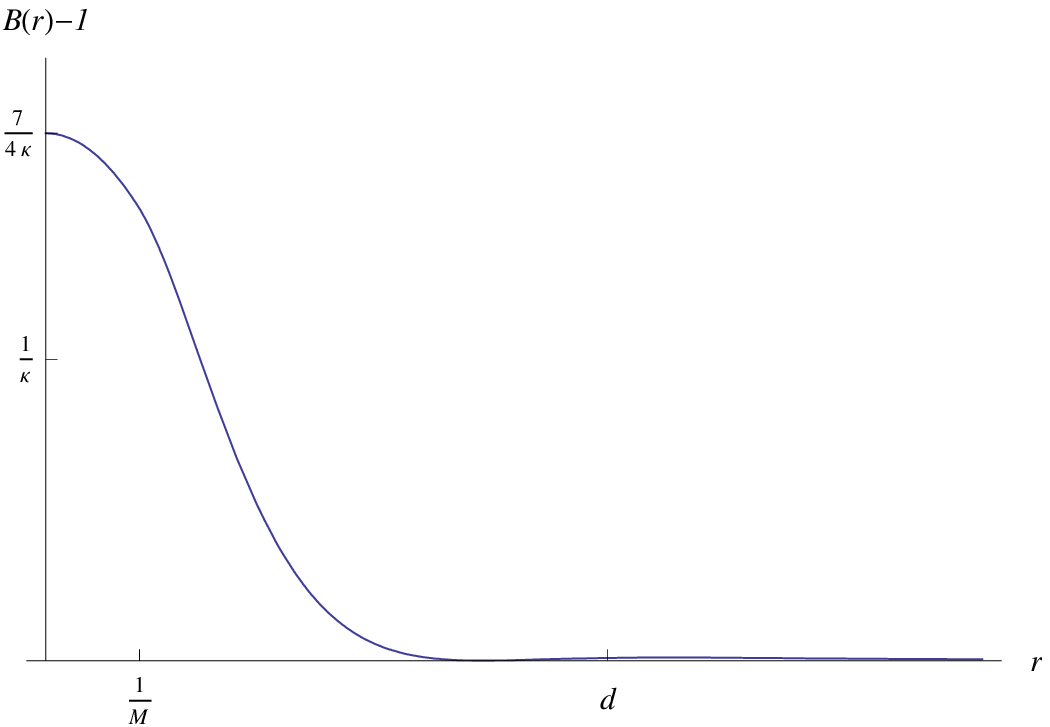,width=.45\textwidth}}\\
\end{center}
\caption{Small $r$ solutions for the scalar field and electric potential}
\label{fig}
\end{figure}


\section{Fermion Dynamics}

In our discussions above we have treated the fermions as a fixed charge background $J_\mu  = J_0 \delta_{\mu 0}$.  We relax this assumption now and introduce dynamics for the fermions via the Thomas-Fermi approximation.  The fermion dynamics are governed by the constant chemical potential $\mu_F$:
\beq
\label{muF}
\mu_F = \sqrt{p_F^2(x)+m_e^2}+e A_0(x) \, .
\eeq
The local number density of fermions is determined by the Fermi momentum: $J_0(x) = p_F^3(x)/(3 \pi^2)$.  In this way the number density of the fermions $J_0$ gets related to the electric potential $A_0$.  For relativistic fermions
\beq
\label{J0}
J_0(x) \simeq  \frac{1}{3 \pi^2} (\mu_F-e A_0(x))^3 \, .
\eeq
The chemical potential gets fixed by the value of the fermion number density in the condensate phase, where $\langle A_0 \rangle = 0$.  If $\bar{J}_0$ represents the number density of fermions in the condensate, then $\mu_F = (3 \pi^2 \bar{J}_0)^{1/3}$.  The photon mass $m_\gamma$ is also defined in terms of $\bar{J}_0$: $m_\gamma \equiv 2e \sqrt{\bar{J_0}/2m_H}$.  In the vortex phase $J_0(x) \rightarrow \bar{J}_0$ for large $x$.

To include the effects of an $x$-dependent $J_0$ into our equations, (\ref{J0}) gets incorporated into the equations of motion (\ref{eom1}).  As a result the equation of motion for $B(x)$ (\ref{eomB}) gets modified.  In the linearized equations, the effect is the addition of a new term for $b(x)$ with a coefficient which scales as $\propto m_H/\mu_F$.  However, it turns out that this new term does not contribute significantly to the solutions.  This is because, in the fourth order differential equation for $b(x)$, terms with coefficient $m_H/\mu_F \propto \kappa^{2/3}$ are subdominant compared to terms with coefficient $\kappa^2$.  Accordingly, the solutions found above in the intermediate and asymptotically large regions are still valid.  It can be shown that the $x \rightarrow 0$ solutions (\ref{A0}), (\ref{B0}), (\ref{F0}) are also unaffected.  In physical terms, the inclusion of the fermion dynamics via the Thomas-Fermi approximation gives rise to ordinary Debye screening.  This screening is subdominant compared to other screening effects in the condensate (see \cite{GGRR3}).  Moreover, the profile of $A_0(x)$ away from the core and within the halo is very shallow, giving rise to a very mild dependence of the charge density on $x$. Hence, the latter can be approximated by a constant, as was done in the previous sections.

The Thomas-Fermi approximation does not capture the possibility of exciting gap-less modes near the Fermi surface.  To include this effect we must calculate the one-loop correction to the gauge boson propagator.  In other words, we must include in the Lagrangian (\ref {lagr0}) the fermion kinetic, mass and chemical potential terms and take into account the known one-loop gauge boson polarization diagram when calculating the gauge boson propagator.  This was done in the second reference 
in \cite{GGRR3} and we use those results here in what follows.

The one-loop correction to the gauge boson propagator gives corrections to the static potential $A_0$.  We are interested in how this correction compares to the potential found in the intermediate region of the vortex (\ref{Blin}).  To estimate its magnitude, we consider a toy model of the vortex.  We find the potential due to a wire of constant linear charge density $\lambda_0$ located at $r=0$.  The linear charge density of this wire is set by the characteristic charge of the vortex: since the scalar charge density varies significantly from its condensate value $e J_0$ at scales $r < 1/M$, it follows that at short distances the linear charge density of the vortex can be approximated by
\beq
\label{lambda0}
\lambda_0 = \frac{e \pi J_0}{M^2} \, .
\eeq
At large distances the vortex is effectively neutral, as mention above.  Thus we expect the one-loop contribution to the static potential to be irrelevant at large scales.

In three dimensions, the charge density of the source is given by
\beq
\label{J0source}
J_0^{\rm source} (r,\theta,z) = \frac{\lambda_0}{\pi r} \delta(r) \, .
\eeq
The static potential is determined from this source and from the $\{00\}$ component of the gauge boson propagator $D_{00}$:
\beq
\label{convol}
A_0(r) = -\int d^3r' \, D_{00}(r-r') J_0^{\rm source}(r') \, .
\eeq
The propagator was found in the second reference in \cite{GGRR3}:  
\beq
\label{D00}
-D_{00}(\omega =0, \k)= \left( \k^2 +m_\gamma^2 +  {4M^4\over \k^2} + F(k^2,k_F,m_f)  \right)^{-1}\,.
\eeq
The function $F(k^2,k_F,m_f)$ is due to the one-loop photon polarization diagram.  It includes both the vacuum and fermion matter contributions.   Here $k_F$ denotes the Fermi momentum and $m_f$ the mass of the fermion.  A complete expression for $F (k^2,k_F,m_f)$ can be found in Ref. \cite {KapustaT}.  We take the expression for $F (k^2,k_F,m_f)$ in the massless ($m_f=0$) limit, which is a good approximation for ultra-relativistic fermions:
\beq
\label{F}
F (\k^2,k_F)=  {e^2 \over 24\pi^2}  
\left (16k^2_F +{k_F (4k_F^2 -3k^2)\over k}\ln ( {2 k_F  +k \over 2 k_F -k })^2
- k^2 \ln ( { k^2 -4 k_F^2\over \mu_0^2} )^2 \right )\, .
\eeq
Here $\mu_0$ stands for the normalization point that appears in the one-loop vacuum polarization diagram calculation.

The function $F$ introduces a shift of the pole in the propagator, corresponding to the``electric mass'' of the photon.  This part of the pole can be incorporated via the Thomas-Fermi approximation, as was done above.  In addition, however, the function $F$ also gives rise to branch cuts in the complex $|\k|$ plane (see Ref.\cite{Walecka} for the list of earlier references on this).  These branch cuts give rise to the additional terms in the static potential which are not exponentially suppressed, but instead have an oscillatory behavior with a power-like decaying envelope.  In a non-relativistic theory they're known as the Friedel oscillations \cite {Walecka}.  In the relativistic theory they were calculated in Refs. \cite {Sivak,KapustaT}.  
We have calculated them in Ref. \cite{GGRR3} 
for the relativistic theory in the presence of the condensate: taking the Fourier 
transform of (\ref{D00}), the dominant contribution due to the branch cuts is
\beq
\label{D00r}
-D_{00}(\bar{r}) = \frac{\alpha_{\rm em}}{\pi^2} \frac{k_F^5 \sin(2k_F\bar{r})}{M^8 \bar{r}^4} \, .
\eeq
Here $\bar{r}$ represents the 3D radius in spherical coordinates, as opposed to the 2D radius $r$.

Using this expression together with expression (\ref{J0source}) in equation (\ref{convol}), the correction to the static potential is
\beq
\label{A0lamdba}
A_0(r) = \frac{\alpha_{\rm em}}{\pi^2} \,\frac{ \lambda_0 k_F^5}{M^8} \,
 \int_{-\infty}^\infty dz' \, \frac{\sin(2k_F\sqrt{z'^2+r^2})}{(z'^2+r^2)^2} \, .
\eeq
An upper bound on the potential can be found by taking $\sin(2k_F\sqrt{z'^2+r^2}) \rightarrow 1$.  After integrating, this gives
\beq
\label{A0lambdaless}
A_0(r) \, < \, \frac{\alpha_{\rm em}}{\pi^2} \,\frac{ \lambda_0 k_F^5}{M^8} \, \frac{\pi}{2 r^3}\, \propto \, \frac{\pi^2}{e^2} \sqrt{\frac{k_F}{m_H}} \frac{1}{m_H^2 r^3} \, .
\eeq

On the vortex solution, the leading term in the potential in the intermediate region is given by expression (\ref{Blin}):
\beq
\label{A0int}
A_0(r) = \frac{m_H}{2e} (B(r)-1) \simeq  \frac{1}{4e\,m_H r^2} \, .
\eeq
Given that both $k_F/m_H \ll1$ and $1/(m_Hr) \ll 1$, we see that the one-loop correction to the potential (\ref{A0lambdaless}) is greatly suppressed compared to the potential found in the vortex solution.  Thus the excitations of the fermions do not significantly alter the vortex solutions.

One further effect that we take into consideration is the Landau quantization of the fermion energy levels due to the presence of the magnetic field in the interior of the vortex.  In the presence of an applied magnetic field, the separation between the Landau  levels is given by  $\omega = eH/m_f$ where $H$ is the magnetic field and $m_f$ is the fermion mass.  Near the core of the vortex where $H \simeq m_\gamma^2/(2e)$ the separation of levels of the fermions is $\omega \simeq m_\gamma^2/m_f$.  Since the photon mass $m_\gamma$ is generally much smaller than the fermion mass, this shift in energy is negligible compared to the typical energy of the fermions.

\section{Energetics and external fields}

We now turn to the question of when it is energetically favorable to form a vortex in the charged condensate.  We start by comparing the average energy density of the vortex to the energy density of the pure condensate.  Above, using Gauss's law, we found that inside the distance $x=1$ the vortex has a small negative charge density implying that in this region the average scalar number density is lower than in the condensate phase.  At distances $x \gg 1$ this charge density is exponentially suppressed indicating that the net charge of the vortex is zero and thus the total average scalar number density is the same in both the vortex phase and the condensate phase.  In calculating the average energy density of the vortex inside the distance $x =1$, we are not interested in the contribution to the energy due to the discrepancy in the number of scalars between the vortex phase and the condensate.  This contribution to the overall difference in energy vanishes at large distances.  Thus we calculate the energy density of the system using ${\cal{H}}'  = {\cal{H}} - \mu_s J_0^{\rm{scalar}}$.  The additional term effectively subtracts off the energy density due to the scalar number density.  We compare ${\cal{H}}'$ in the vortex phase to ${\cal{H}}'$ in the condensate.

The Hamiltonian density ${\cal{H}}'$ can be calculated from the Lagrangian ${\cal L}'$ (\ref{lagr}):
\beq
\label{ham}
{\cal{H}}' =  \tfrac{1}{2}H^2 + \tfrac{1}{2}E^2 +\tfrac{1}{2}(2eA_0+\mu_s-\dot{\alpha})^2 \sigma^2-\mu_s(2eA_0+\mu_s-\dot{\alpha})\sigma^2  \, ,
\eeq
We have simplified the Hamiltonian using the equations of motion (\ref{eom1}), (\ref{eom2}) and have taken boundary terms to be negligible.  The magnetic field $H$ and the electric field $E$ are defined as usual
\beq
\label{em}
H= \frac{1}{r} \frac{d}{dr} (r A_\theta)\, , ~~~~E = -\frac{dA_0}{dr} \, .
\eeq
The fourth term in the Hamiltonian is exactly $- \mu_s J_0^{\rm{scalar}}$ as we would expect.  The third term is due to the energy of the scalar field.  Unlike in the abelian Higgs model, the energy density of the scalar field in the center of the vortex, i.e. in the ``normal" phase, is lower than in the condensate phase.  However, the energy density of the vortex is still greater than that of the condensate alone, due to the gradients of the scalar field and due to the intermediate region $1/\sqrt{\kappa} \ll x \ll 1$ in which the values of both the potential and the scalar field are greater than their condensate values.  This contribution to the energy density is roughly equal in magnitude to the contributions coming from the electric and magnetic fields.  On the condensate solution, the Hamiltonian density is identically zero: ${\cal{H}}'_{\rm CC} = 0$.

For large $x$, deviations away from the condensate are exponentially 
suppressed and thus differences in energy between the two phases are negligible. 
So to find the average energy density within the vortex, we integrate the Hamiltonian 
density over an area of radius $x=1$ and then divide by the total area.  
The average energy density within the radius $x \le 1$ is
\beq
\label{eps}
\epsilon_{\rm{ave}} = \int^1_0 2x\,dx\,{\cal{H}}' \, .
\eeq
The Hamiltonian density can be further simplified using the equations of motion 
(see Ref. \cite{SJTS} for more details).  In terms of the dimensionless 
variables defined above equation (\ref{eps}) becomes:
\beq
\label{eps2}
\epsilon_{\rm{ave}} =\frac{1}{2} m_H J_0  \int^1_0 x\,dx\,(B(x)-1) \, (3-F(x)^2) \, .
\eeq
For the region $x < 1/\sqrt{\kappa}$ we use solutions (\ref{B0}) and (\ref{F0}) and in the intermediate region $1/\sqrt{\kappa} < x <1 $ we use solutions (\ref{Blin}) and (\ref{Flin}) with the coefficients found from matching (\ref{bfco}).  Upon integration, the average energy density is
\beq
\label{eps3}
\epsilon _{\rm{ave}}= \frac{m_H J_0}{4 \kappa^2} \left(\log{\kappa}+ 14 \right) \,.
\eeq
The numerical coefficients should not be taken too literally given the approximations made in obtaining the solutions which yield the above result.  However, the overall scaling of the energy density $\epsilon _{\rm{ave}} \propto m_H J_0 (\log{\kappa})/\kappa^2$ is remarkably independent of the matching radius and other details of the solutions.  As the energy density of the condensate is effectively zero (${\cal{H}}'_{\rm CC} = 0$), the above expression represents the difference in energy between the two phases.  

To see when it is energetically favorable for the condensate to form vortices, we now consider placing the condensate in an external field $H_{\rm ext}$ pointed along the $z$-axis.  We shall see that the magnetic properties of the charged condensate resemble those of a superconductor.  In particular, when $\kappa \gg 1$, the charged condensate resembles a type II superconductor.  When an external magnetic field $H_{\rm ext}$ is applied to the condensate, below a critical value $H_{c1}$ surface currents oppose the penetration of the field and the induction $B_{\rm ind}$ is zero in the bulk of the condensate.  For $H_{\rm ext} > H_{c1}$ magnetic flux penetrates the condensate in the form of vortices.  At another critical value of the magnetic field $H_{c2}$ the normal phase is restored and the induction $B_{\rm ind}$ is equal to the applied field $H_{\rm ext}$.  In what follows we determine the critical values of the fields $H_{c1}$ and $H_{c2}$.

Given the energy density $\epsilon$ of the vortex phase above the pure condensate phase, we can find the value of the magnetic field $H_{c1}$ at which it becomes energetically favorable to form vortices.  In the absence of an external field, it is never energetically favorable to form vortices as the energy density of a vortex is greater than that of the pure condensate.  In the presence of a small external magnetic field, below $H_{c1}$, the condensate must expel the magnetic field entirely from its bulk in order to remain in the condensate phase.  This requires energy; the energy per volume needed to expel the external field is $\tfrac{1}{2} H^2_{\rm ext}$.  If vortices form in the condensate then the energy required to expel the magnetic field is smaller than if the field were to be completely expelled.  More specifically, if the vortices give rise to an average magnetic field in the condensate $B_{\rm ind}$, then the energy needed to expel the remaining magnetic field would be $\tfrac{1}{2} (H_{\rm ext}-B_{\rm ind})^2$.  The energy gained by forming vortices is the difference between this energy and the energy required to expel the magnetic field entirely.  Assuming that $B_{\rm ind}$ is small compared to $H_{\rm ext}$ near the transition point, this difference can be approximated by $\tfrac{1}{2} (2 B_{\rm ind} H_{\rm ext})$.  Thus for a high enough external field, the energy $\epsilon$ lost in creating a vortex is compensated by the energy gained in expelling a smaller magnetic field $B_{\rm ind} H_{\rm ext}$.  In order for formation to be energetically possible, we must have $\epsilon \leq B_{\rm ind} H_{\rm ext}$.  The equality determines the critical external field $H_{c1}$.  (See Ref. \cite{Abri}.)

Suppose the number of vortices per area in the condensate is given by $N$.  Then the energy density due to the formation of vortices is given by $\epsilon = N \lambda$ where $\lambda$ is the energy density per unit length of a vortex.  Using $\epsilon _{\rm{ave}}$ found above (\ref{eps2}) as the energy density of a single vortex,
\beq
\label{lambda}
\lambda = \frac{\pi}{m_\gamma^2} \epsilon _{\rm{ave}} \, .
\eeq
The induction $B_{\rm ind}$ is given by
\beq
\label{Bind}
B_{\rm ind} = N \, \oint {\bf{A \cdot}} \d {\bf{l}} = \frac{2 \pi N}{2 e} \, .
\eeq
Combining these expressions, the critical field $H_{c1} = N \lambda/(B_{\rm ind})$ is given by
\beq
\label{Hc1}
H_{c1} = \frac{m_\gamma^2}{8e} \left(\log(\kappa)+ 14 \right) \, .
\eeq
The final expression for $H_{c1}$ is independent of the number density of vortices $N$.  It follows that if it is energetically favorable to create one vortex, then it will be even more energetically favorable to create many, up to the point than interactions between vortices become significant.  At distances greater than $r = 1/m_\gamma$ we expect fields outside the vortices to be exponentially suppressed and thus the vortices to be effectively non-interacting.  So at the transition point $H_{c1}$, it is likely that the number density of vortices is of the order $N \simeq m_\gamma^2/\pi$.

If we take $J_0 \simeq (0.15-0.5 {\rm ~MeV})^3$, a reasonable value for white dwarfs, 
this gives a magnetic field of roughly $H_{c1} \simeq (10^{7}- 10^{9})$ Gauss.  
Thus, the vortex lines should be expected to  be present in the bulk 
of the helium-core white dwarf stars  with strong enough magnetic fields.  

A sufficiently high magnetic field will disrupt the condensate entirely.  One way to approximate the magnetic field at which this transition occurs is to consider the density of vortices in high external magnetic fields.  When the cores of the vortices begin to overlap, then the scalars are mostly returned to the normal phase.  We define the core of the vortex to correspond to $x=1/\sqrt{\kappa}$, or equivalently, $r=1/M$ as this is the region in which the VEV of the scalar field drops to zero.  As $N$ is the number of vortices per area, at the transition point $N \simeq M^2/\pi$.  We define the critical external field at this point to be $H_{c2}$.  When the condensate enters the normal phase, the induction $B_{\rm ind}$ is equal to the external magnetic field $H_{c2}= B_{\rm ind} = 2 \pi N/(2e)$.  From these two expressions we find $H_{c2}$:
\beq
\label{Hc2}
H_{c2} = \frac{M^2}{e} \, .
\eeq
For $J_0 \simeq (0.15 - 0.5 {\rm ~MeV})^3$, $H_{c2} \simeq (10^{13}-10^{15})$ Gauss.  
This is well above the values of the fields expected to be present in a majority 
of white dwarf stars.  Thus the external magnetic field is unlikely to be 
large enough to push the condensate into the normal phase.

We note that both $H_{c1}$ and $H_{c2}$ given above were determined at zero temperature.  Generally, we expect these expressions (\ref{Hc1}), (\ref{Hc2}) to be valid at temperatures well below the condensation temperature.  

Finally, in type II superconductors the dependence of the critical temperature 
on the magnetic field is well-approximated by $T^{\prime 2}_c/T_c^2 \simeq (H_c-H)/H_c$,
where $T^{\prime }_c$ is the transition temperature  when the magnetic field 
$H$ is present. We expect a similar relation to be valid in our case too. 
Hence, as long as the value of the magnetic field is not too 
close to either critical value,  the change of the transition temperature due 
to the magnetic field should be  small.  Near the critical values, however, the change 
of the phase transition temperatures  (from the normal to the vortex phase and from 
the vortex phase to the phase with no magnetic field) could change significantly.
The would be crystallization temperature will also change, and the charged condensation 
may or may not be favorable for close-to-critical magnetic fields\footnote{For 
discussions of other magnetic field effects in very highly magnetized white dwarfs, 
see \cite {Single}}.

\section{The low energy effective Lagrangian}
For a system of helium-4 nuclei and electrons, one can consider a non-relativistic effective Lagrangian for the order parameter $\Phi$, as the helium nuclei are non-relativistic in the condensate phase.  This Lagrangian is less restrictive than the relativistic Lagrangian in that it is not required to be Lorentz invariant.  The non-relativistic Lagrangian must give rise  to the Schr\"{o}dinger equation for the order parameter in lowest order in the fields and it must respect the appropriate symmetries of the physical system, including translational, rotational and Galilean symmetries as well as gauge invariance.  (See \cite{GGRR3,GGDP} for more details.)  Such a Lagrangian was initially proposed by Greiter, Wilczek and Witten in the context of superconductivity \cite{GWW}.  We used it here to describe the charged condensate:
\beq
\label{Leff}
{\cal L}_{eff} = {\cal P} \left ( {i\over 2} ( \Phi^*  D_0 \Phi -  (D_0 \Phi)^* \Phi)-
{| D_j  \Phi|^2  \over 2m_H} \right )\,,
\eeq
where $D_0 \equiv (\partial_0  - 2ie A_0)$,  $D_j \equiv ( \partial_j - 2ie A_j) $, and ${\cal P}(x)$ is a polynomial function of its argument.  We could introduce a chemical potential for the scalars into the argument of ${\cal P}(x)$ in the form $\mu_{NR} \Phi^* \Phi$.  The relationship between the relativistic chemical potential $\mu_s$ and the non-relativistic chemical potential is given by $\mu_s = m_H +\mu_{NR}$.  Thus the neutral condensate where $\langle A_0 \rangle = 0$ and $\mu_s =m_H$ corresponds to $\mu_{NR} = 0$ (assuming that the 
quartic and other interactions are neglected).  Again, one could also include a quartic term 
$\lambda (\Phi^* \Phi)^2$.  However, as long as $m_H^3 \gg \lambda J_0$ and 
$\lambda \lesssim 1$, this term can be neglected\footnote{The chemical potential and 
quartic terms would have to be retained if we were  to discuss temperatures near 
the phase transition point.}.

In the condensate where the VEV of $\Phi$ is nonzero, we can express $\Phi$ in term of a modulus and phase: $\Phi = \Sigma\, {\rm exp}(i\Gamma)$.  Written in terms of fields $\Sigma$ and $\Gamma$, the effective Lagrangian (\ref {Leff}) takes the following form: 
\beq
\label{Leff1}
{\cal L}_{eff} = {\cal P} \left ((2e A_0-\partial_0 \Gamma) \Sigma^2 - \frac{1}{2 m_H} (\nabla_j  \Sigma )^2 -\frac{1}{2 m_H}(2e A_j-\partial_j \Gamma)^2 \Sigma^2  \right )\,.
\eeq
The gauge field couples to the electron density as $-eA_0 J_0$, thus we include this term in the Lagrangian (\ref{Leff1}).  Once again we work in the unitary gauge and set $\Gamma = 0$.  The equations of motion which follow from (\ref{Leff1}) then have the following static solution:
\beq 
\label{sol}
2e\Sigma^2 =e J_0\,,~~~ A_\mu=0,~~~{\cal P}^{\prime}(0)=1\,.
\eeq 
The above solution describes a neutral system in which the helium-4 charge density $2e\Sigma^2$ exactly cancels the electron charge density $-eJ_0$.  Since on the solution the argument of (\ref {Leff1}) is zero, the condition ${\cal P}^{\prime}(0)=1$ is satisfied by any polynomial function for which the first coefficient is normalized to one: ${\cal P} (x) = x + C_2 x^2+...$.

The condensate solution sets a preferred Lorentz frame.  We consider the dynamics of small perturbations above this background.  We express $\Sigma$ in terms of a perturbation $\tau$ above the condensate value:
\beq
\label{Sigmapert}
\Sigma(x) = \sqrt{m_H} \left(\sqrt{\frac{J_0}{2 m_H}}+\tau(x) \right) \, .
\eeq
The Lagrangian, including the gauge field kinetic term, expanded to second order in fields becomes:
\beq
\label{Lnrpert}
{\cal L}_{eff} = -{1 \over 4} F_{\mu\nu}^2-{1 \over 2} (\partial_j \tau)^2+{1 \over 2}C_2 m_H J_0 m_\gamma^2 A_0^2 - {1\over 2} m_\gamma^2 A_j^2+2m_H m_\gamma A_0 \tau \, .
\eeq
We can compare this Lagrangian to the one obtained from the relativistic theory (\ref{lagr}) 
by enforcing Lorentz invariance, i.e. by demanding that the ``electric" mass of the gauge 
field be equal to the ``magnetic" mass.  This would fix the value of $C_2 = 1/(m_H J_0)$.  
With this value of $C_2$, the Lagrangian for small perturbations above the condensate in 
the non-relativistic theory is identical to the Lagrangian for small perturbations 
in the relativistic theory found in Ref. \cite{GGRR1}, up to a time derivative for $\tau$.

However, we do not in general expect that the low energy effective theory will obey the 
Lorentz invariant condition $C_2 =1/(m_H J_0)$.  Instead, $C_2$ must be fixed by the 
particular physics of the system.  It's worth noting that even if $C_2$ were to be 
fixed via Lorentz invariance, introducing fermion dynamics via the Thomas-Fermi 
approximation introduces an additional term into the Lagrangian (\ref{Lnrpert}) 
of the form $e^2 \mu_F^{2/3} A_0^2/\pi^2$ which breaks the degeneracy between 
the electric and magnetic masses.  This is a typical scale by which we'd expect  the 
electric and magnetic masses squares to differ from each other. 

In section 3, in order to find vortex solutions in the intermediate region $1/\sqrt{\kappa} \ll x \ll 1$ and the asymptotic  region $x \gg 1$, we treated the electric potential $A_0$ and the scalar field $\sigma$ in the linear approximation, but kept higher order terms for the vector potential.  Thus to determine the applicability of the solutions found above to the non-relativistic effective theory, we should consider higher order terms in $A_j$ than the ones given in (\ref{Lnrpert}).  We also restore the phase $\Gamma$.  Given ${\cal P} (x) = x + C_2 x^2+...$, the equations of motion to next-to-leading-order are
\begin{align}
\label{eomeffB}
-\partial^{\mu} F_{\mu 0} &= 2e\left[1+2 C_2 \Sigma^2 (2eA_0-\partial_0 \Gamma)\right]\Sigma^2 -e J_0 \, , \\
\label{eomeffA}
-\partial^{\mu} F_{\mu j}& = 2e (2eA_j-\partial_j \Gamma) \Sigma^2 \, , \\
\label{eomeffF}
-\nabla^2 \Sigma &= \left[2 m_H (2eA_0-\partial_0 \Gamma) + 4 C_2 m_H \Sigma^2(2eA_0-\partial_0 \Gamma)^2 \right] \Sigma-(2eA_j)^2 \Sigma \, .
\end{align}
If we take $\Sigma = \sqrt{m_H} \sigma$ and $C_2 =1/(m_H J_0)$, then the first two 
equations of motion above (\ref{eomeffB}), (\ref{eomeffA}) are the same as in the non-relativistic case (\ref{eom1}), up to second order in small fields.  The third equation (\ref{eomeffF}) has an extra factor of $(2eA_0)^2 \Sigma$ compared to equation (\ref{eom2}).  However, since this term is second order in $A_0$ and we treated $A_0$ in the linear approximation, this does not alter our solutions in the intermediate and asymptotic regions.  Thus for $C_2 =1/(m_H J_0)$, 
the vortex solutions found above for $x \gg 1/\sqrt{\kappa}$ are also solutions 
for the non-relativistic effective theory. As  we mentioned above, however, 
we would expect the realistic value of $C_2 $ to be different from the one we 
used by the quantity  $ (e^2 \mu_F^{2/3}/m_H J_0 m_\gamma^2) $. However, 
as we have shown in section 4, such corrections are subdominant  because the 
value of the photon electric mass is smaller that the value set by the scale $M$.

As in the relativistic case, the solutions formally break down near $x=1/\sqrt{\kappa}$ 
when the change in the scalar field becomes of order $1$ and thus the linear approximation 
is no longer valid.  More realistically though, we do not expect the effective field theory 
to hold  at distances shorter than the interparticle separation $x \propto 1/\kappa^{1/3}$. 
Instead it will cease to be a valid description of the physics before reaching 
$x=1/\sqrt{\kappa}$.

\vspace{0.5cm}

We can use the non-relativistic formalism to consider the effects of the rotation 
of a white dwarf star on the magnetic field in its interior\footnote{We thank Daniel Stein for raising this issue.}.
In this formalism the scalar number density and current density are given respectively by:
\beq
\label{scalcur}
J_0 ^{\rm{scalar}} = \Phi^* \Phi \, , ~~~~~~ J_j ^{\rm{scalar}} = \frac{-i}{2m_H}\left[(D_j\Phi)^* \Phi-\Phi^*(D_j \Phi) \right] \, .
\eeq 
The number density is related to the current density by $J_j ^{\rm{scalar}} = J_0 ^{\rm{scalar}} v_j$ where $v_j$ is the velocity vector of the rotating scalar particles.   Using the change of variables defined above, $\Phi = \Sigma\, {\rm exp}(i\Gamma)$, we can use these expressions to find $v_j$:
\beq
\label{vj}
v_j = \frac{1}{m_H} \, (2eA_j-\partial_j \Gamma) \, .
\eeq
This known result is notably different from that of a superfluid in which the scalar field does not couple to a gauge field.  In the absence of the $A_j$ term in the above expression, one would conclude that ${\bf \nabla} \times {\bf v} = 0$ and thus the scalar condensate does not support rotation.  Instead, in the presence of the gauge field we find
\beq
\label{rotv}
{\bf \nabla} \times {\bf v} = \frac{2e}{m_H} \, {\bf H} \, .
\eeq 
The magnetic field ${\bf H}$ is called the London field \cite{London}.

The velocity vector ${\bf v}$ can be written in term of the angular velocity ${\bf v} = {\bf \Omega} \times {\bf r}$.  It follows that, for constant ${\bf \Omega}$, the rotation of ${\bf v}$ is given by ${\bf \nabla} \times {\bf v} = 2 {\bf \Omega}$.  Accordingly, the magnetic field can be expressed in terms of the angular velocity:
\beq
\label{HLondon}
{\bf H} = \frac{2m_H}{2 e} \, {\bf \Omega} =  \frac{2eJ_0}{m_\gamma^2} \, {\bf \Omega} \, .
\eeq
Here $J_0$ is the fermionic number density.  Thus the condensate of helium-4 nuclei can rotate with the rest of the star, unlike a neutral condensate.  The consequence is a small, constant magnetic field in the bulk of the condensate.

Varying the Lagrangian (\ref{Leff1}) with respect to $A_j$ gives $2 e J_j ^{\rm{scalar}} = e J_j$, where $J_j$ is the fermion current density.  Using $J_j = J_0 v_j$, it follows that the fermion velocity vector is equal to the scalar velocity vector.  The electrons and the helium-4 nuclei rotate together in the core of the star.  At the surface however, there is a thin layer of helium-4 nuclei that is slightly out of rotation with the rest of the star.  This feature becomes evident upon finite volume regularization of the system.  The thickness of the layer is roughly $1/m_\gamma$.  This surface layer is what gives rise to the London field in the interior of the star \cite{Sauls}.  To  estimate the  value of the London field we take the 
angular velocity of a helium white dwarf star to be  $\Omega \sim  10^{-2}$ Hz.  
The resulting London field is $H \simeq 10^{-6}$ Gauss.  This field is present even in 
the absence of vortices.  However, it is too small to affect any of the results 
given above.

\vspace{0.2in}

\begin{center}
{\bf   Acknowledgments}
\end{center}

We'd like to thank Malvin Ruderman, Daniel Stein  and  Alexander 
Vilenkin  for useful discussions and correspondence.  GG  was 
supported by  the NSF and NASA grants PHY-0758032, NNGG05GH34G. RAR 
acknowledges the NYU James Arthur graduate fellowship support.

\end{document}